\documentclass[aps,prl,twocolumn,amsmath,superscriptaddress]{revtex4}

\usepackage{amssymb}
\usepackage{amsmath}
\usepackage{epsfig}

\begin{document}

\title{Public Channel Cryptography: Chaos Synchronization and Hilbert's Tenth Problem}

\author{Ido Kanter}
\affiliation{Department of Physics, Bar-Ilan University,
Ramat-Gan, 52900 Israel}
\author{Evi Kopelowitz}
\affiliation{Department of Physics, Bar-Ilan University,
Ramat-Gan, 52900 Israel}
\author{Wolfgang Kinzel}
\affiliation{Institute for Theoretical Physics, University of
W\"urzburg, Am Hubland, 97074 W\"urzburg, Germany}

\begin{abstract}
The synchronization process of two mutually delayed coupled
deterministic chaotic maps is demonstrated both analytically and
numerically. The synchronization is preserved when the mutually
transmitted signal is concealed by two commutative private filters
that are placed on each end of the communication channel. We
demonstrate that when the transmitted signal is a convolution of
the truncated time delayed output signals or some powers of the
delayed output signals synchronization is still maintained. The
task of a passive attacker is mapped onto Hilbert's tenth problem,
solving a set of nonlinear Diophantine equations, which was proven
to be in the class of NP-Complete problems. This bridge between
two different disciplines, synchronization in nonlinear dynamical
processes and the realm of the NPC problems, opens a horizon for a
new type of secure public-channel protocols.

\end{abstract}


\maketitle


Chaotic systems are very unpredictable and two chaotic systems,
starting from almost identical initial states, end in completely
uncorrelated trajectories\cite{1}. Nevertheless, two chaotic
systems which are coupled by some of their internal variables may
synchronize to a common identical chaotic motion\cite{2,3}.
Unpredictability\cite{abel} or chaos synchronization, of coupled
chaotic systems, have attracted a lot of attention, mainly because
of the potential to build a secure communication protocol based on
artificial chaotic systems\cite{3,Einat1} or coupled chaotic
lasers\cite{lasers6,nature,lasers2_MCPF}.

The security of a public-key encryption protocol based on chaos
synchronization relies on the fact that two chaotic systems, $A$
and $B$, synchronize by bi-directional interaction whereas a third
unit $E$, which is only driven by the transmitted signal cannot
synchronize. However, it is not obvious that this is possible at
all. On one hand, the two mutually coupled chaotic systems
influence the dynamics of each other and can accelerate the
synchronization by enhancing coherent moves, whereas the
unidirectionally coupled system, an attacker, cannot influence the
synchronization process. On the other hand, the attacker is
allowed to record and to manipulate his recorded signals, without
affecting the synchronization process\cite{MRZ,shamir}. Note that
the two partners, $A$ and $B$, are not allowed to exchange any
secret information; the attacker $E$ knows all the details which
$A$ knows about the system of $B$ and vice versa.

For identical partners which synchronize by a bi-directional
signal we recently presented a proof that an attacking unit
coupled unidirectionally can synchronize as well\cite{proof}. The
proof is valid for any type of transmitted signals, for instance,
a nonlinear function of the time delayed output signals. For
non-identical partners which can synchronize, using for instance
private commutative filters, it may be difficult for the attacker
to synchronize and to reveal the time dependent output signal of
the parties\cite{proof}, but one cannot exclude efficient advanced
software or hardware attacks. A hardware attacker consists of a
similar chaotic setup to those of the synchronized chaotic
partners, whereas a software attacker is able to mathematically
manipulate the recorded signal.

In order to exclude any possible software advanced attack, we map
the task of the attacker onto one of the NP-Complete (NPC)
problems\cite{npc}. The NPC problems are the most difficult
problems in NP (non-deterministic polynomial time) and at present,
all known deterministic algorithms for NPC problems require
running time that is exponential with some tunable parameters of
the problem. The main goal of this Letter is to bridge between two
different disciplines, synchronization in nonlinear dynamics and
the realm  of the NPC problems. The establishment of such a bridge
proves the lack of any possible efficient software attack, while
the mutually coupled chaotic partners are synchronized. Note that
the definition of the known NPC problems is static\cite{npc}, and
here we map a dynamical process onto an NPC problem.

Hilbert's tenth problem is the tenth on the list of Hilbert's
problems of $1900$\cite{hilbert}. Its statement is as follows;
given a set of Diophantine equations, polynomials with integer
coefficients, finding an integer solution that satisfies the set.
The solution of a general set of Diophantine equations is known to
be undecidable\cite{diophantine1,papa,gurari}. However, some
subsets of the Diophantine equations are known to be decidable and
belong to the class of NPC problems\cite{gurari,diophantine1}. A
class of Hilbert's tenth problem is to find an integer solution of
the following set of Diophantine equations\cite{gurari}
\begin{equation}
\label{dioph} D\vec{y}= \vec{\sigma}(z),
\end{equation}
where $D$ is an $ m \times n$ matrix of rational constants,
$\vec{y}=(y_1,~...,y_n)$ and
$\vec{\sigma}=(\sigma_1(z),~...,\sigma_m(z))$ is a column vector.
The $\{\sigma_i(z)\}$ are polynomials with a finite degree greater
than one. Finding a non negative integer solution
$(y_1,~...,y_n,z)$ to the above set was proven to belong to the
class of NPC problems\cite{gurari}. In this Letter we map the task
of an attacker in the scenario of two synchronizing chaotic units
onto this NPC problem.

We start by defining our synchronization process of two
interacting units. Consider two iterated chaotic maps $x^A$ and
$x^B$, which their dynamics are controlled by a general
self-feedback function $S_f$ and a general coupling function $S_c$
which are both nonlinear functions of the history $\tau$ steps
back
\begin{equation}
\begin{split}
x_t^A=S_f(\vec{x}^A_t) + S_c(\vec{x}^B_t)\\ x^B_t=S_f(\vec{x}^B_t) +
S_c(\vec{x}^A_t)
\end{split}
\label{eqn_1}
\end{equation}
where $\vec{x}_t=(x_{t-1},..,x_{t-\tau})$.

\begin{figure}[h]
\includegraphics[width=0.235\textwidth,height=0.22\textwidth]{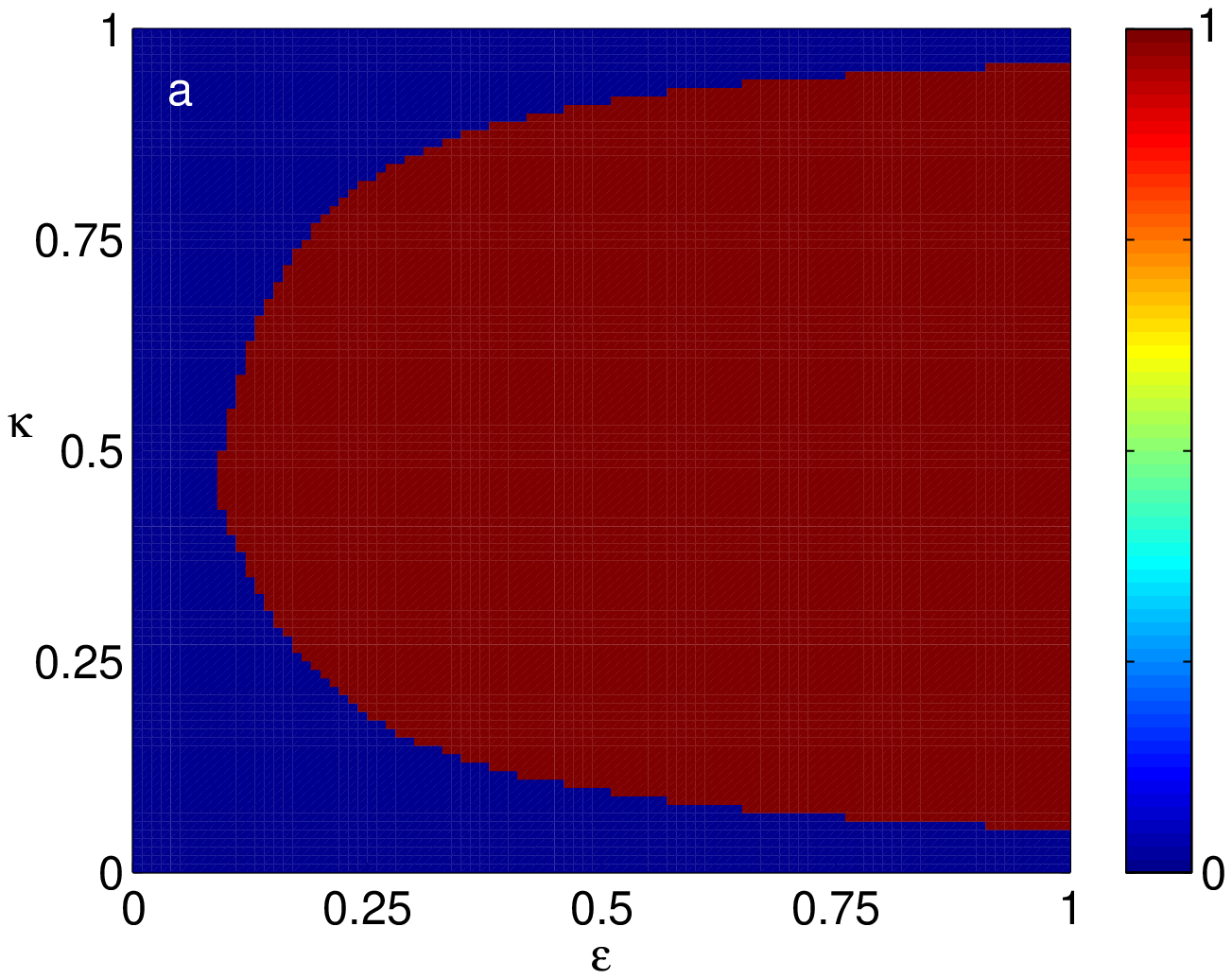}
\includegraphics[width=0.235\textwidth,height=0.22\textwidth]{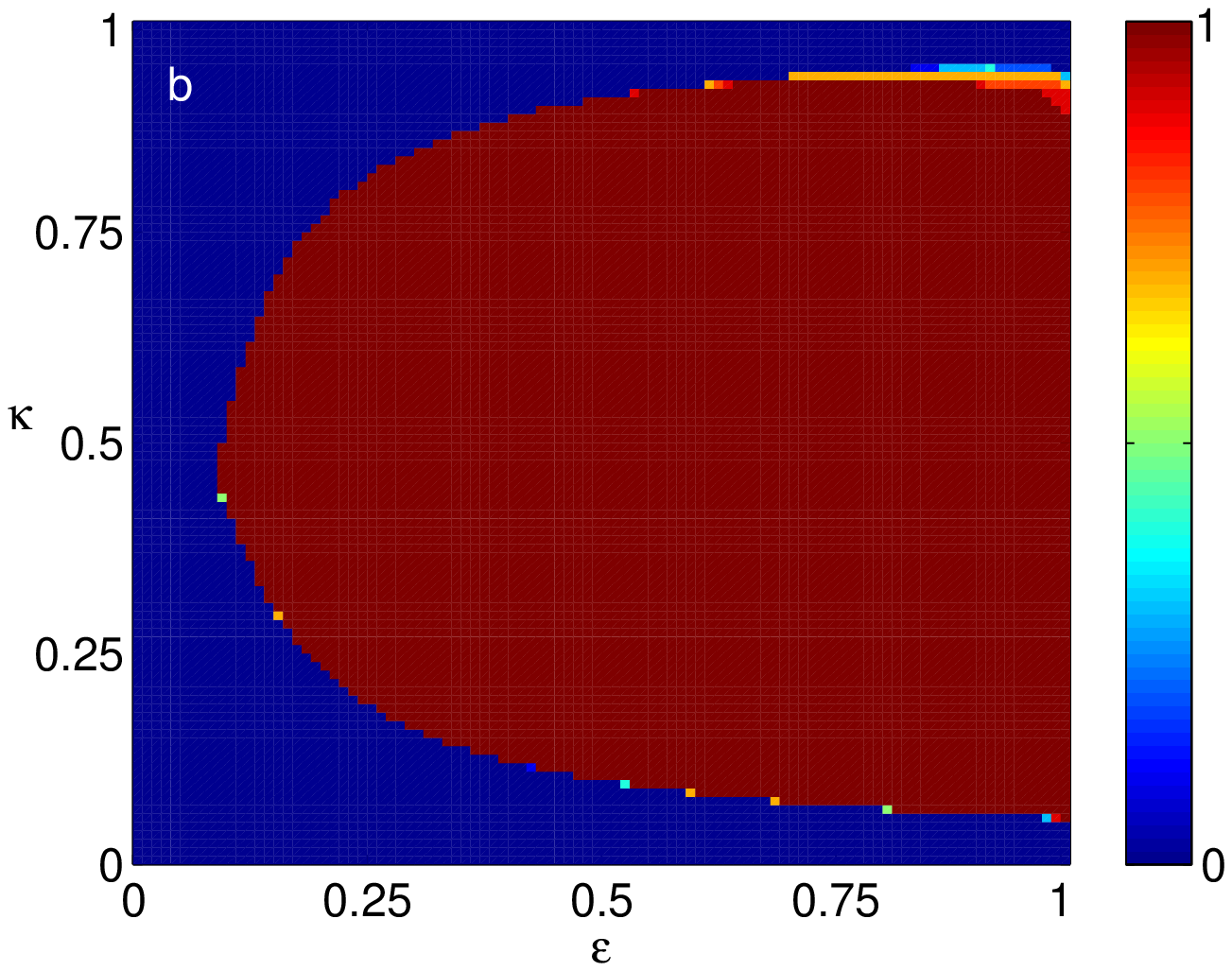}
\caption{\label{fil_1} Semi-analytic results for the fraction of the
phase space, $(\varepsilon,\kappa)$, where synchronization is
achieved for a Bernoulli map with $\tau=100$ and $a=1.1$. (a) With
the absence of filters, synchronization is achieved only in the red
regime. (b) The probability to synchronize in the case of un-clipped
filters with $N=10$ and $\phi=2$.} \vspace{-0.6cm}
\end{figure}

Do the two mutually coupled chaotic maps synchronize under such
circumstances? The positive answer is demonstrated below for the
simplest chaotic maps, the Bernoulli map\cite{2}. The dynamics of
the two mutually coupled units $x_t^A$ and $x_t^B$ can be analyzed
analytically and is given by
\begin{equation}
\label{eqn_x}
\begin{split}
x_{t}^A=(1-\varepsilon)f(x_{t-1}^A)+\varepsilon[\kappa
f(x_{t-\tau}^A)+(1-\kappa)R^A(\vec{x}^B_t)]\\
x_{t}^B=(1-\varepsilon)f(x_{t-1}^B)+\varepsilon[\kappa
f(x_{t-\tau}^B)+(1-\kappa)R^B(\vec{x}^A_t)]
\end{split}
\end{equation}
where $f(x)=(ax)\mod1$, and a Bernoulli map is chaotic for
$a>1$\cite{joha}. The parameter $\varepsilon$ indicates the weight
of the delayed terms, $\kappa$ stands for the strength of the
self-coupling term, and $R^{A,B}(\vec{x}^{B,A}_t)$ are the received
signals of each partner. Note that $[0,1]$ is the allowed range for
$\varepsilon$ and $\kappa$. For the simple case of
$R^{A,B}(\vec{x}^{B,A}_{t})=f^{A,B}(x^{B,A}_{t-\tau})$, a linear
expansion of the distance $d_t=x_t^A-x_t^B$ leads to
$d_t=(1-\varepsilon)ad_{t-1}+\varepsilon
a(2\kappa-1)d_{t-\tau}$\cite{physica_d,joha}. By assuming that the
distance converges/diverges exponentially in time, $d_t \propto
c^t$, we find that the largest conditional Lyapunov exponent is
negative and synchronization is achieved for
$(a-1)/2a\varepsilon<\kappa<(2a\varepsilon+1-a)/2a\varepsilon$ as is
depicted in figure \ref{fil_1}(a).

In order to map the task of an attacker on this synchronization
process to the presented NPC problem, we have to include the
following four adjustments to the system: (a) private commutative
filters, (b) transmission of integer signals, (c) additional
nonlinear terms to the transmitted signal and (d) periods of
cutoffs in communication. Our next goal is to explain each one of
these adjustments and to show that synchronization is still
maintained when applying all of the adjustments simultaneously,
and finally to show that the task of the attacker is mapped onto
the NPC problem, eq.~(\ref{dioph}).

\begin{figure}[h]
\includegraphics[scale=0.6]{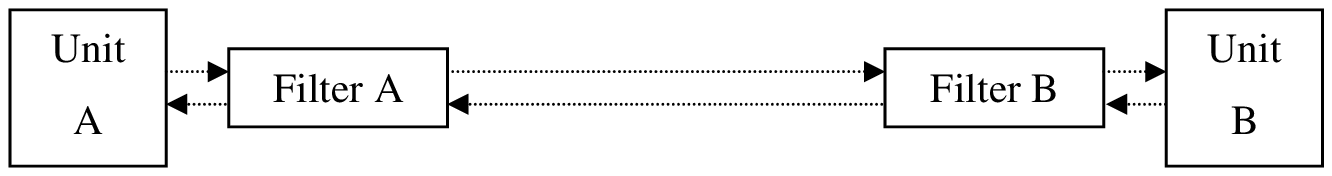}
\caption{\label{fil_2} A setup of two time-delayed mutually
coupled units, where each unit has a filter influencing both
transmitted and received signals.}\vspace{-0.3cm}
\end{figure}

The first adjustment is extending the configuration,
equation~(\ref{eqn_1}), to the case of non-identical units $x^A$ and
$x^B$. Both units are using different functions (filters) $g_A$ and
$g_B$, and the two transmitted signals are $g_A(\vec{x}^A_t)$ and
$g_B(\vec{x}^B_t)$, see figure \ref{fil_2}. These functions are
private, only $x^A$ knows $g_A$ and $x^B$ knows $g_B$. The coupling
functions $S_c(\vec{x}^B_t)$, $S_c(\vec{x}^A_t)$ are simply the
received signals which are $g_A(g_B(x^B_t))$ and $g_B(g_A(x^A_t))$,
respectively. In order to preserve synchronization as a fix point of
the dynamics we only use filters that commute, $g_A(g_B(\vec{x})) =
g_B(g_A(\vec{x}))$. Since an attacker does not know the filters he
cannot use them for his hardware attack.

The most simple {\it commutative} filter one can consider is
convolution. The transmitted signal is defined by
\begin{eqnarray}
\label{Ttabb} T_t^{A,B} = g_{A,B}(\vec{x}_t^{A,B})
=\sum_{\nu=0}^{N-1}{K_{A,B}^{\nu}f(x^{A,B}_{t-\nu})}
\end{eqnarray}
where $K_A^{\nu},K_B^{\nu}\in[0,1]$ are the private keys (filters)
chosen randomly by each one of the partners and
$\nu=0,1,\ldots,N-1$. We demand that
$\sum_{\nu=0}^{N-1}{K_{A,B}^{\nu}}=1$, in order to ensure that the
convolved signal is limited by $[0,1]$.

Before arriving at the other end of the channel, the transmitted
signal $T$ encounters the second filter. Therefore, the received
signal for units $A$ and $B$ is
\begin{eqnarray}
\label{Ttab2} R_t^{A,B} = g_{A,B}(\vec{T}_t^{B,A})
=\sum_{\mu,\nu=0}^{N-1}{K_B^{\nu}K_A^{\mu}f(x_{t-\nu-\mu}^{B,A})~.}
\end{eqnarray}

\begin{figure}[h]
\includegraphics[width=0.235\textwidth,height=0.22\textwidth]{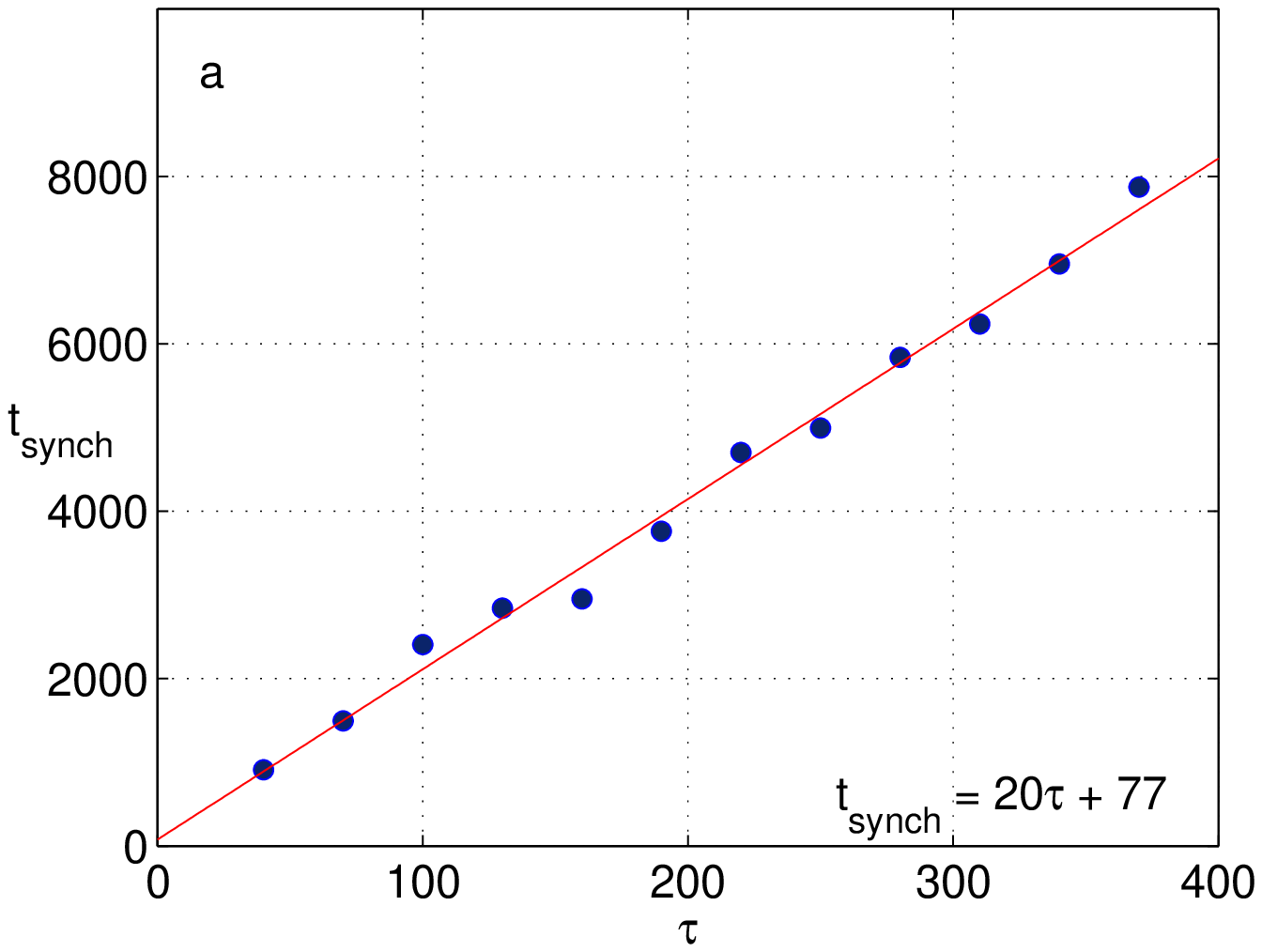}
\includegraphics[width=0.235\textwidth,height=0.22\textwidth]{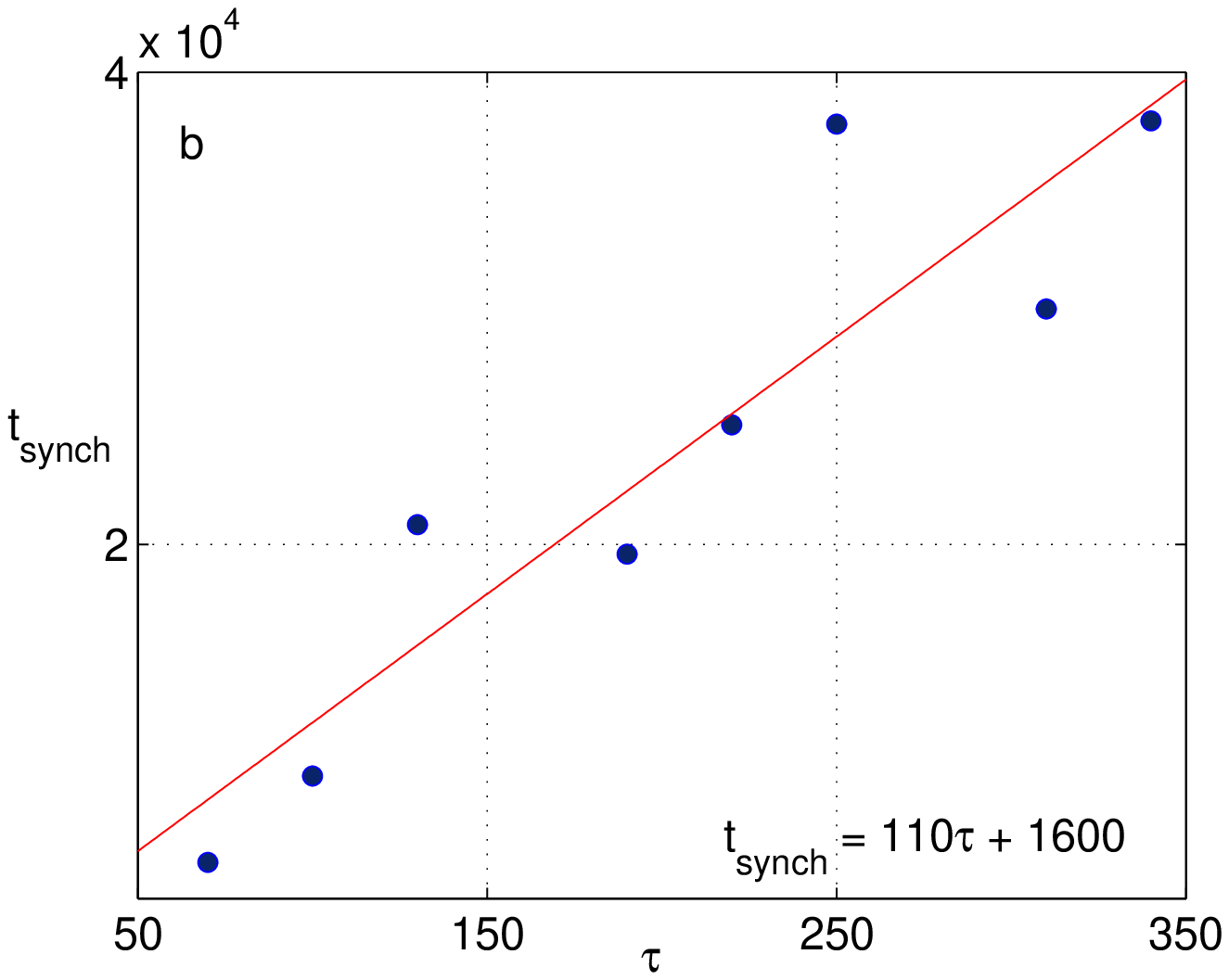}
\caption{\label{fil_3} Simulation results for the synchronization
time, $t_{synch}$, as a function of $\tau$ for $a=1.1$, $N=10$ and
$\phi=2$:
(a) linear filters, and (b) with quantization
$m=6$, and an additional quantized nonlinear term. $\rho_t=2,3,4,5$
with equal probability, $C_t\in[0,0.1]$, $N_0=40( t\mod 40)-5$,
$N_1=20$ and $N_2=20$, the solid lines were obtained by linear
fitting.} \vspace{-0.5cm}
\end{figure}

We measure the synchronization time $t_{synch}$ as a function of
$\tau$ and found that in order to achieve linear synchronization
time for $N\gg 1$, the strengths of the filter coefficients, the
keys, have to follow a power-law $K_A^\nu,K_B^\nu\propto\frac{
\xi^\nu_{A,B}}{(1+\nu)^\phi}$, where $\xi^\nu_{A,B}$ is a random
number between $[0,1]$. Figure \ref{fil_3}(a) exemplifies the
linear scaling of $t_{synch}(\tau)$ for $N=10$ and $\phi=2$. The
synchronization phase space was analyzed semi-analytically by
assuming that the distance between the partners converges/diverges
exponentially with time and then solving the characteristic
polynomial and the largest eigenvalue
numerically\cite{physica_d,joha}. Since the values of the private
keys $K_A,K_B$ are random, we calculate the probability of
achieving synchronization in the phase space of
$(\varepsilon,\kappa)$ using sampling of random sets of keys. In
figure \ref{fil_2} we compare the semi-analytic results for the
regimes of synchronization for the basic setup without filters (a)
and with filters (b). We found that even in this case, the regime
of synchronization is almost unchanged,

The next two adjustments ((b) and (c)) to the synchronization
process is modifying the transmitted signal to be composed of
clipped output keys and signals, and also to include a nonlinear
term of the past output signal. Practically, the precision of the
computer is $m_0$ decimal digits, and the key-filters and output
signals consist of only $m \ll m_0$ most significate decimal
digits (or integers after multiplying by $10^m$). Adopting these
two adjustments the transmitted signal has the following form:
\begin{eqnarray}
\label{Ttabb} T_t^{A,B}
=\sum_{\nu=0}^{N-1}{K_{A,B}^{\nu}f(x^{A,B}_{t-\nu})}+
C_{t}[f(x^{A,B}_{N_0})]^{\rho_t}
\end{eqnarray}
where $K_{A,B}$ are the clipped keys, and $f(x^{A,B}_{t-\nu})$ are
the clipped output signals. $C_{t}[f(x^{A,B}_{N_0})]^{\rho_t}$ is
the non-linear term which is not convolved in the current filters,
$C_t$, $\rho_t$ and $N_0$ are public constants used simultaneously
by both partners. $C_t \in[0,1]$ and is also clipped, the power
$\rho_t$ is an integer and
$N_0 (<t-N)$ is a time step from the past. Since the partners are
using different private keys (filters), synchronization is a fix
point of the dynamics only when each partner subtracts his own
nonlinear term before applying the convolution using his key.
Therefore, the received signal in case of synchronization is
\begin{eqnarray}
\label{Ttab2} R_t^{A,B} =
g_{A,B}(\vec{T}_t^{B,A}-C_{t}[f(x^{A,B}_{N_0})]^{\rho_{t}})
\\\nonumber
=\sum_{\mu,\nu=0}^{N-1}{K_B^{\nu}K_A^{\mu}f(x_{t-\nu-\mu}^{B,A})}
\end{eqnarray}
It is clear that synchronization is a fixed point of the dynamical
process, since after the convolution at the receiver the nonlinear
terms appear only in the form
$C_{t}[f(x^{B}_{N_0})^{\rho_t}-f(x^{A}_{N_0})^{\rho_t}]$ which
vanishes when the partners are synchronized. It is worthy to note
that since the keys are normalized and $C_t>0$ it is possible that
the received signal is greater than one, however in practice it
does not affect the synchronization process, and alternatively one
can apply mod 1 again on the received signal. Both methods give
the same regime of synchronization.

\begin{figure}[h]
\includegraphics[width=0.235\textwidth,height=0.22\textwidth]{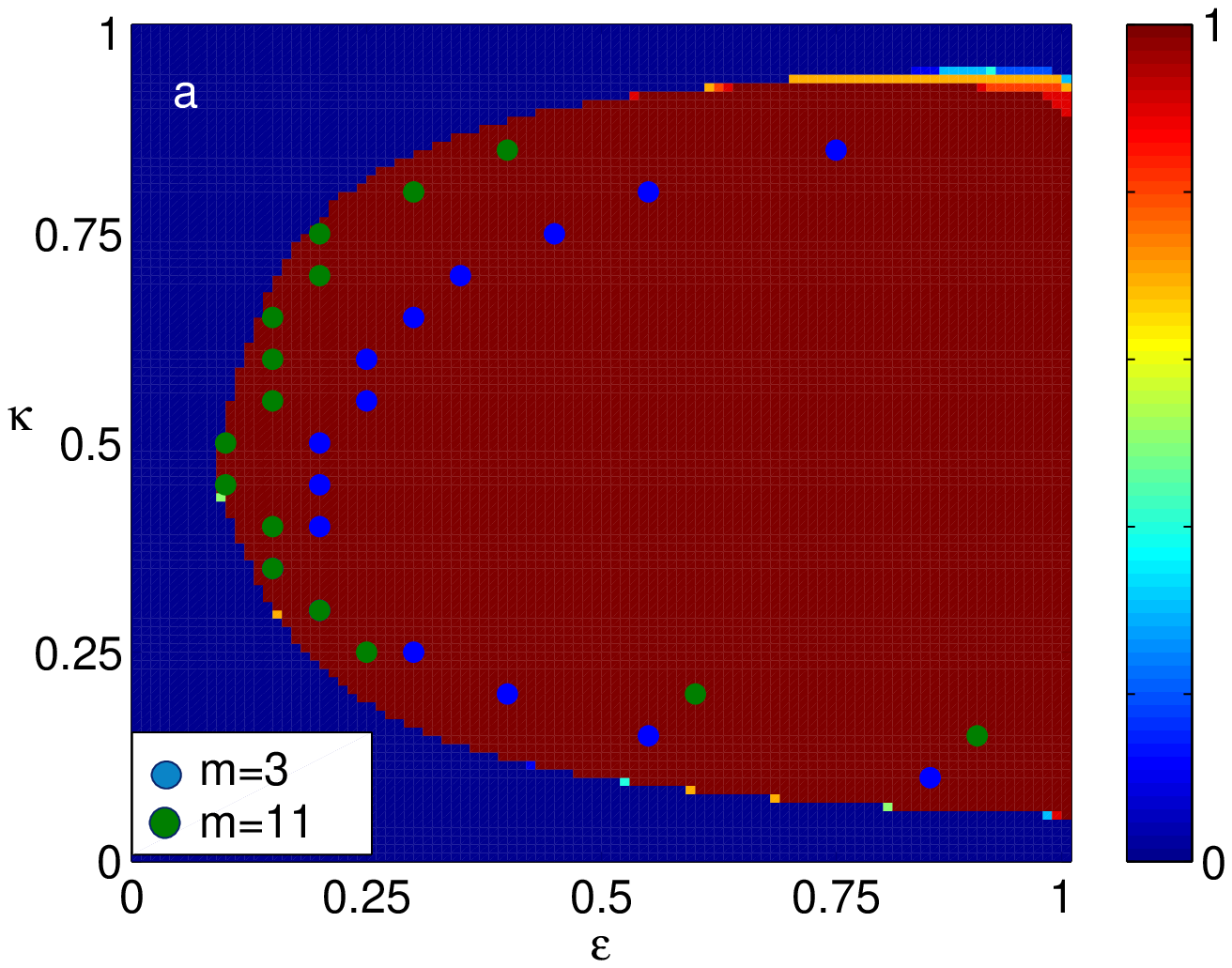}
\includegraphics[width=0.235\textwidth,height=0.22\textwidth]{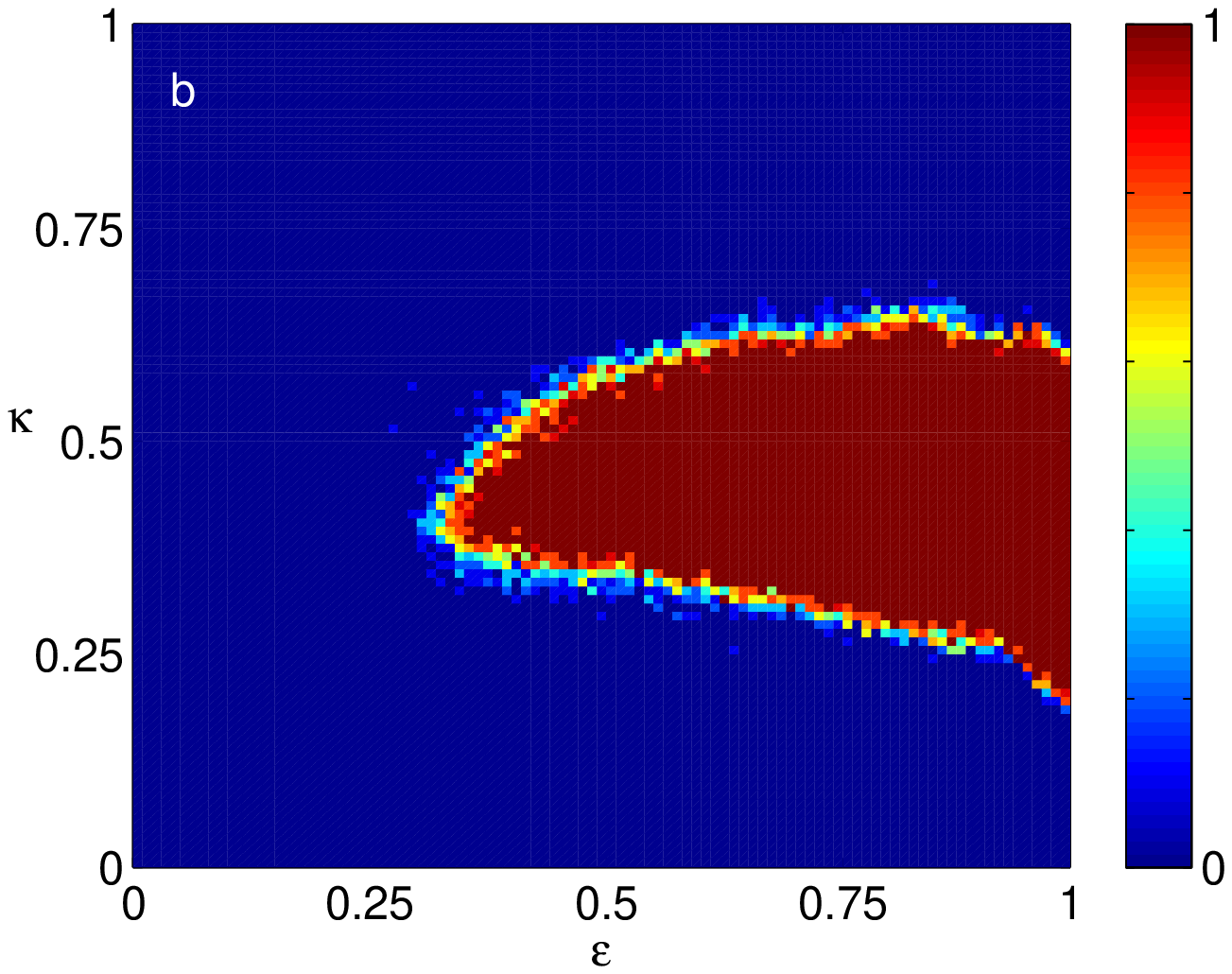}
\caption{\label{fil_4} Simulation results for the fraction of the
phase space, $(\varepsilon,\kappa)$, where synchronization is
achieved for $\tau=100$, $a=1.1$, $N=10$ and $\phi=2$. (a) With
quantized linear filters for $m=3,11$ (red-unclipped) (b) With
quantization $m=6$ and the same parameters as for Fig. 3(b).}
\end{figure}

For the case of clipped keys and output signals simulations with
$m_0=32$ indicate that the regime in the phase space where
synchronization exists is only slightly affected by the
quantization of the keys and the transmitted signals. A typical
result for different values of $m$ is depicted in figure
\ref{fil_4}(a).

The last adjustment ((d)) of our setup is the implementation of
dynamical filters. For $N_1$ steps the partners are using the above
mentioned prescription. For the next $N_2$ steps no communication
between the partners occurs, and each partner is updating his states
following his own history of continuous signals with $\kappa=1$ in
eq.~(\ref{eqn_x}). After each period of silence, $N_2$, each partner
is selecting a new set of private filters, and in addition, they
select the nonlinear contribution to the transmitted signal to be a
function of the signal at a time step, $N_0$, belonging to the
previous silence period\cite{ido1}.

Simulations indicate that while the synchronization time and phase
space are affected by the nonlinear additional term in
eq.~(\ref{Ttabb}) and by the silence periods, $t_{synch}$ still
scales linearly with $\tau$ as depicted in figure \ref{fil_3}(b),
and synchronization is achieved in a non-negligible fraction of the
phase space. For instance, synchronization for $\rho_t=2,3,4,5$ with
equal probability, $C_t\in[0,0.1]$, $N=10$, $N_1=20$, $N_2=20$ and
$N_0=40( t\mod 40)-5$ is depicted in figure \ref{fil_4}(b).

We now turn to discuss the complexity of a unidirectional
listener. To avoid any software attack or any other advanced
attacks we now map the task of the attacker to the NPC problem,
eq.~(\ref{dioph}). Assuming a synchronization state, $\vec{x}_t^A
= \vec{x}_t^B \equiv \vec{x}_t$. In one time step, the transmitted
signals on both directions, $T_t^{A,B}$, consist of $3N-2$ unknown
variables: $\{K^{\nu}_{A,B}\},~f(x_{t}),~...,~f(x_{t-N+1})$. On
the next time step, two new equations emerge: $T_{t+1}^{A,B}$.
These equations consist of previously unknown variables and one
new unknown variable $f(x_{t+1})$. Therefore by adding more time
steps we are adding more equations than new variables. Actually
the number of required equations to decode the keys of length $N$
is $6(N-1)$. Therefore, the number of  required iterations is
$3(N-1)$. In order for a passive attacker to construct the entire
signal, he needs to eavesdrop over at least $3(N-1)$ successive
time steps. His task in such a scenario is therefore to solve a
set of nonlinear Diophantine equations\cite{diophantine1,papa}.
The nonlinearity emerges since the attacker does not know neither
the integer keys, $K_{A,B}^{\nu}$, nor the history of the clipped
output signals of the partners.

In order to map our synchronization problem to the proven NPC
problem, eq. (1) we choose $N_1$ to be in the range of $N<
N_1<3(N-1)$ (see for instance fig.~\ref{fil_4}(b)). Hence, the task
of the attacker is to find the complete set of solutions for the
nonlinear Diophantine equations (unknown clipped keys and history of
clipped signals), and next to find the correct solution for the
observed dynamical synchronization process. The number of solutions
is at least one, but can be unbounded, hence, the complexity of the
attacker is at least NPC, where the complexity of the problem
increases with $N$. The silence regime, $N_2>N$ was selected to
guarantee that the set of Diophantine equations the attacker has to
solve consists of nonlinear terms of only one past clipped output
signal (as formally required by eq.~(\ref{dioph})). Note that  the
use of time-dependent filters eliminates, in the jargon of nonlinear
dynamics, eliminates any approximated reconstruction of the
trajectory based on Takens embedding theorem\cite{20} since the
transmitted signal is a discontinuous function of the chaotic
variables.

Note that also with the lack of adjustment (c) (the nonlinear term
in eq.~(\ref{Ttabb}) the problem reduces to the solvability of
linear Diophantine equations which belongs to the class of
NPC\cite{papa,gurari,fiege}. However, finding a solution of a set
of linear Diophantine equations may be feasible in practice, in
polynomial time using heuristic or probabilistic methods
\cite{frenkel}.

We prove semi-analytically that the security of the simplest
synchronization process (Bernoulli map) consists of $\tau$ ${\it
time-independent}$ local Lyapunov exponents. In simulations we
obtained similar results also for more structured maps and for the
Lang-Kobayashi differential equations governing the behavior of
semiconductor lasers. Note that transmitted signal in lasers is
quantized by the number of photons and in principle convolutional
filters can be implemented.

We thank Johannes, Kesstler, Uri Feige and Aviezri Fraenkel for many
fruitful discussions.


\vspace{-0.4cm}
\begin{thebibliography}{9}
\vspace{-0.4cm}

\bibitem{1}
H. G. Schuster, W. Just. \emph{Deterministic Chaos}. Wiley VCH,
(2005).

\bibitem{2}
A.~Pikovsky, M.~Rosenblum, J.~Kurths. \emph{Synchronization: A
Universal Concept in Nonlinear Sciences}, Cambridge Univ. Press,
N.Y. (2001).

\bibitem{3}
L. M.~Pecora, T. L.~Carroll, Phys. Rev. Lett. {\bf 64}, 821
(1990).

\bibitem{abel}
A. Abel and W. Scharz, Proc. IEEE, {\bf 90}, 691 (2002).


\bibitem{Einat1}
E.~Klein, R. Mislovaty, I.~Kanter, W.~Kinzel, Phus. Rev. E {\bf
72}, 016214 (2005).

\bibitem{lasers6}
G. D. VanWiggeren, R. Roy, Science {\bf 279}, 1198 (1998).

\bibitem{nature}
A. Argyris, D. Syvridis, L. Larger, V. Annovazzi-Lodi, P. Colet,
I. Fischer, J. Garcia-Ojalvo, C. R. Mirasso, L. Pesquera, K. A.
Shore, Nature {\bf 438}, 343 (2005).

\bibitem{lasers2_MCPF}
E. Klein, N. Gross, E. Kopelowitz, M. Rosenbluh, W. Kinzel, L.
Khaykovich, I. Kanter, Phys. Rev. E {\bf 74}, 046201 (2006).


\bibitem{MRZ}
M.~Rosen-Zvi, E.~Klein, I.~Kanter and~W. Kinzel, Phys. Rev. E.
{\bf 66}, 066135 (2002).

\bibitem{shamir} A. Klimov, A. Mityagin and A. Shamir,
ASIACRYPT 288-298 (2002).

\bibitem{proof} I. Kanter, E. Kopelowitz, W. Kinzel and J. Kestler,
arXiv:0712.2712v1.

\bibitem{npc} M. R. Garey and D. S. Johnson, \emph{Computers and
Intractability}, W H Freeman  Corporation, (1979).

\bibitem{hilbert}
Davis,~M. Amer. Math. Monthly 80, 233-269 (1973).



\bibitem{diophantine1}
http://mathworld.wolfram.com/DiophantineEquation.html, and
references threrin.

\bibitem{papa} C. H. Papadimitriou
\emph{Computational Complexity}, Addison Wesley (1994).



\bibitem{gurari} E. M. Gurari and O. H. Ibarra, J. Assoc.Comput. Mach.
{\bf 26}, 567-581 (1979).
\bibitem{joha}
J.~Kestler, W.~Kinzel, I.~Kanter, Phys. Rev. E {\bf 76}, 035202
(2007).


\bibitem{physica_d}
S.~Lerpi, G.~Giacomelli, A.~Politi, F. T.~Arecchi, Physica D {\bf
70}, 235 (1993).

\bibitem{ido1} Synchronization under a scenario of scilence periods is
based on the fact that resycnhronization time of two mutually
coupled chaotic maps is shorter in comparison to desynchronization
time. This inequality was also recently observed in an experiment
of two mutually coupled semiconductor lasers, Phys. Rev. Lett. 98,
154101 (2007).

\bibitem{20} F. Takens, in Dynamical Systems and Turbulence (War-
wick 1980), edited by D. A. Rand and L.-S. Young (Springer-Verlag,
Berlin, 1980), vol. 898 of Lecture Notes in Mathematics, pp.
366-381.



\bibitem{fiege} C. H.~Papadimitriou, JACM 28, 4 (1981) proves that this
problem is in NP. A proof that this problem is in NPC was offered by
Uri Feige based on the following reduction from SAT. For every
variable $z$ in the input SAT formula, introduce two variables in
the system of equations, $z_0$ for a positive literal corresponding
to the variable, $z_1$ for a negative literal and add the equation
$z_0 + z_1 = 3$. This constraint forces exactly one of these
literals to have value $1$ (interpreted as "true") and the other to
have value $2$ (interpreted as false). For every clause $C_i$ add
the equation stating that the sum of the literals in the clause plus
a new variable $c_i$ is equal to twice the number of variables in
the clause. This can be satisfied by setting $c_i$ to be equal to
the number of satisfied literals in the clause, which is positive if
and only if the formula is satisfiable.

\bibitem{frenkel} I. Borosh and A. S. Fraenkel, Math. Comp.
{\bf 20}, 107 (1966).




\end{thebibliography}
\end{document}